\documentclass[twocolumn, oneside]{article}

\usepackage{graphicx} 

\usepackage{geometry}
\mathcode"65"65 
\usepackage{upgreek} 
\usepackage{nicefrac} 
\usepackage{amsmath}
\usepackage{cite}     
\usepackage[hyphens]{url} 
\usepackage[hidelinks]{hyperref}

\usepackage{color} \usepackage{varwidth}
\usepackage{soul}
\setulcolor{red}
\setul{0em}{1.5pt}

\newcommand{\redline}[1]{#1} 

\usepackage{authblk} 

\begin{document}
\twocolumn[   \begin{@twocolumnfalse}

\title{Extracting losses from asymmetric resonances in micro-ring resonators}

\author{O.~Reshef}
\author{M.~G.~Moebius}
\author{E.~Mazur}

\affil{John A. Paulson School of Engineering and Applied Sciences, Harvard
University, 9 Oxford Street, Cambridge, Massachusetts 02138, USA\vspace{-2em}}

\date{} 

\maketitle

\begin{abstract}
\noindent Propagation losses in micro-ring resonator waveguides can
be determined from the shape of individual resonances in their transmission
spectrum. The losses are typically extracted by fitting these resonances
to an idealized model that is derived using scattering theory. Reflections
caused by waveguide boundaries or stitching errors, however, cause
the resonances to become asymmetric, resulting in poor fits and unreliable
propagation loss coefficients. We derive a model that takes reflections
into account and, by performing full-wave simulations, we show that
this model accurately describes the asymmetric resonances that result
from purely linear effects, yielding accurate propagation loss coefficients.
\vspace{2em}
\end{abstract}

\end{@twocolumnfalse} ]

\section{Introduction}

Ring resonators are one of the simplest and most commonly used components
in photonic integrated circuits. Their high quality factors and ease
of fabrication in any photonic platform make them useful for applications
in wavelength-selective filters and multiplexers~\cite{Little1997a,Barwicz2004},
optical delay lines~\cite{Boyd2006}, switches and modulators~\cite{Almeida2004,Dong2007},
and in other nonlinear applications enabled by the resonantly-enhanced
intensity build-up~\cite{E.Heebner2002,Foster2006,Levy2009,Kippenberg2011,Okawachi2011,Hausmann2014}.
Additionally, due to their simple analytic transfer function, ring
resonators are also used for fabrication characterization; the individual
resonances in the transmission spectrum of a ring can be measured
and fit to extract propagation losses~\cite{McKinnon2009,Reshef}.

This fitting method has been shown to yield accurate propagation losses
for ideal devices~\cite{McKinnon2009}. However, experimental concerns
in real devices often cause resonances to exhibit distortions. This in turn makes it so
the transfer function fails to capture all of the physics of the system, thus
yielding poor fits. Doing so generates unreliable values for propagation losses,
making it more difficult to assess fabrication quality.

There are many reasons why the transfer function of micro-ring resonators might exhibit asymmetries, the most common type of distortion that can reduce the reliability of the fitting process. For example, nonlinear interactions, such as by optical bistability, are known to cause asymmetries~\cite{Heebner1999,Almeida2004a,Bogaerts2012}.
In terms of linear devices, roughness-induced coherent back-scattering or \redline{polarization rotation} within the ring has been shown to generate significant frequency domain distortions as well as resonance-splitting~\cite{Gorodetsky2000, Morichetti2006, Zhang2008a, Morichetti2010, Ballesteros2011, Ji2016, Li2016a}.
\redline{Another possible source for asymmetric resonances are fabrication imperfections in the directional coupler of the ring~\mbox{\cite{Cusmai2005}}.}

Aside from these considerations \redline{that originate within the ring itself}, back-reflections from \emph{outside} of the ring, which are
common in realistic devices, can also yield asymmetries in the resonances that are linear in
origin~\cite{Fan2002, Choy2012,Hu2013a,Lukyanchuk2010,Liang2006, Li2011}. For example, polarization mixing or fabrication defects in the waveguide leading to and from
the resonator can cause partial reflections that yield asymmetric
Fano-like resonances in the otherwise symmetric spectrum of a ring
resonator~\cite{Morichetti2006, Liang2006, Dong2007,Choy2012,Fan2002}. These asymmetries
are most prevalent in smaller, low-loss systems with high quality
factors, and pose an increasing challenge as fabrication quality further
improves.

A significant amount of work has already been done on coupled-cavity
formalisms, even in the context of integrated circuits~\cite{Zhang2008a, Strain2009, Hu2013a, Wu2014, Li2016a}.
Here, we treat the specific case for extracting propagation losses from a
micro-ring resonator.
We generalize the familiar transfer function for micro-ring
resonators to develop a model that includes the interference caused
by accidental reflections in waveguides. Using full-wave simulations
we demonstrate that these asymmetries are caused by reflections and
are independent of input power. We also discuss situations in which
reflections become significant. We use this generalized transfer function
to extract the propagation loss of the simulated devices, and demonstrate
that the model remains reliable in the presence of strong reflections,
unlike the standard symmetric model. This generalized transfer function
is important in the experimental characterization of micro-ring resonators
with inherent reflections.

\section{Theory}

\subsection{Standard symmetric transfer function}

\begin{figure*}[htbp]
\centering
\includegraphics[width=1\linewidth]{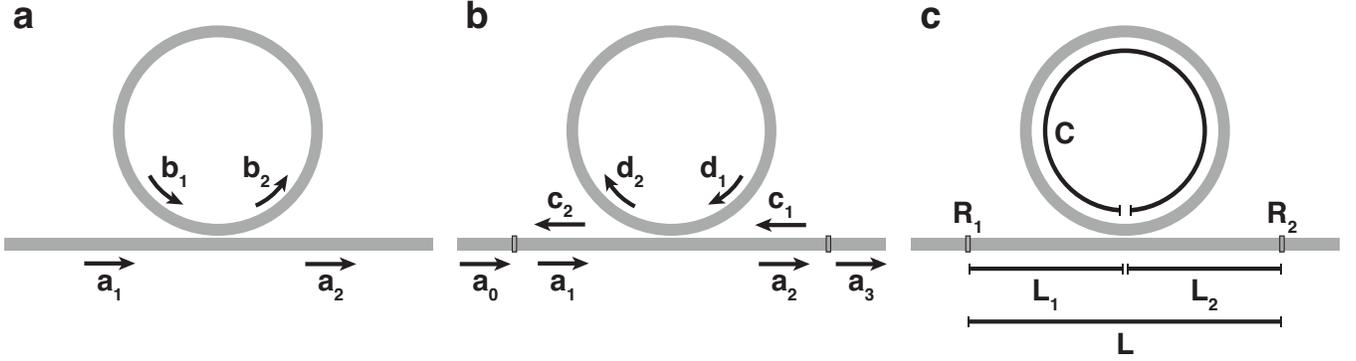}
\caption{\textbf{a}) The right-traveling field in the waveguide excites a counter-clockwise-propagating mode $b_{2}$ and another right-traveling mode $a_{2}$. These two modes interfere destructively when operating on resonance. \textbf{b)}~Partial reflectors at the input and output of the waveguide enable left-traveling fields and clockwise-propagating modes. These fields form a secondary resonance (\emph{i.e.,} $a_{1}\rightarrow a_{2}\rightarrow c_{1}\rightarrow c_{2}\rightarrow a_{1})$ that interferes with the whispering gallery modes in the ring to produce asymmetries in the combined spectrum. \textbf{c)}~The simulated device: a ring resonator of circumference $C$ coupled to a bus waveguide of length $L$ between a pair of reflectors $R_{1}$ and $R_{2}$.}
\label{Fig: ring_schematic}
\end{figure*}

The transmission of a micro-ring resonator coupled to a single waveguide
(\emph{i.e.,} an input port and no add or through port) has been well
characterized and can be derived using scattering theory~\cite{McKinnon2009,Bogaerts2012,Yariv2000,Heebner2004a,Yariv1991a}.
We begin by describing the set of interactions that yield the standard
symmetric transfer function. The complete set of electric fields is
illustrated in Fig.~\ref{Fig: ring_schematic}. The input and output
fields are denoted by $a_{1}$ and $a_{2}$, respectively. The input
field $a_{1}$ is incident on a directional coupler which excites
a counter-clockwise-propagating mode $b_{2}$:
\begin{equation}
b_{2}=\sqrt{1-t^{2}}a_{1},\label{eq:b2_temp}
\end{equation}
where $t$ is the transmission (or self-coupling) coefficient that
corresponds to the field that remains in the waveguide after traversing
the coupling region.

While propagating through the circumference $C$ of the ring, the
field $b_{2}$ accumulates a phase $\phi=\omega C/c=2\pi n_{\textrm{eff}}C/\lambda$
and experiences a propagation loss $\alpha_{\textrm{z}}$ due to scattering
and fabrication imperfections, yielding a field $b_{1}$:
\begin{eqnarray}
b_{1} & = & e^{-\alpha_{z}C/2}e^{i\phi}b_{2}.
\end{eqnarray}
The total propagation loss in the ring is denoted by $\alpha$:
\begin{equation}
\alpha^{2}\equiv e^{-\alpha_{z}C},\label{eq:alpha_loss}
\end{equation}
so that $\alpha\rightarrow1$ in the lossless case and $\alpha\rightarrow0$
as the propagation loss approaches infinity. \redline{The loss coefficient $\alpha$ represents all of the loss mechanisms in the system, including contributions from waveguide scattering, bending losses and the excess insertion losses of the directional coupler.}

Part of the field $b_{1}$ couples back into $b_{2}$. Adding this
contribution to Eq.~\ref{eq:b2_temp}, we obtain:
\begin{eqnarray}
b_{2} & = & \sqrt{1-t^{2}}a_{1}+tb_{1}\nonumber \\
 & = & \sqrt{1-t^{2}}a_{1}+t\alpha e^{i\phi}b_{2}.\label{eq:b2}
\end{eqnarray}

The total output field $a_{2}$ is the sum of the field that is directly
transmitted from $a_{1}$ through the coupler and the field that couples
back from the ring from $b_{1}$:
\begin{eqnarray}
a_{2} & = & ta_{1}-\sqrt{1-t^{2}}b_{1}\nonumber \\
 & = & ta_{1}-\sqrt{1-t^{2}}\alpha e^{i\phi}b_{2}\label{eq:a2}
\end{eqnarray}

The transmission for the ring resonator $T_{\textrm{RR}}(\phi)$ is
the solution of the coupled Eqs.~\ref{eq:b2}--\ref{eq:a2}:
\begin{eqnarray}
T_{\textrm{RR}}(\phi) & \equiv & \left|\frac{a_{2}}{a_{1}}\right|^{2}\nonumber \\
 & = & \left|\frac{t-\alpha e^{i\phi}}{1-\alpha te^{i\phi}}\right|^{2}\nonumber \\
 & = & \frac{t^{2}+\alpha^{2}-2\alpha t\cos\phi}{1+\alpha^{2}t^{2}-2\alpha t\cos\phi}.\label{eq:symmetric_ring_expanded}
\end{eqnarray}

Both parameters $\alpha$ and $t$ are dimensionless and range from
0 to 1. $T_{\textrm{RR}}(\phi)$ is a symmetric function of $\phi$,
which only appears in the argument of a cosine. The transfer function
also remains symmetric upon interchanging $\alpha$ and $t$, as is
seen more evidently in the expanded Eq.~\ref{eq:symmetric_ring_expanded}.
As a consequence, the two coefficients contribute similarly to the
transfer function and thus are often difficult to distinguish when fitting~\cite{McKinnon2009,Rasoloniaina2014a}.
They can be disentangled by fitting to resonances measured from multiple
devices or to a range of resonances from a single device because the
propagation loss is expected to remain constant as a function of wavelength
or other device parameters (\emph{e.g.,} coupling gap, ring radius), \redline{whereas the coupling coefficient should vary~\mbox{\cite{McKinnon2009, Delage2009}}}.

\subsection{Asymmetric transfer function}

We now add a pair of partial reflectors with reflectivities $R_{1}$
and $R_{2}$ to the waveguide in locations that surround the ring
at distances $L_1$ and $L_2$ away, respectively (Fig.~\ref{Fig: ring_schematic}c).
The additional fields that are introduced
are illustrated in Fig.~\ref{Fig: ring_schematic}b.

The original output $a_{2}$ now excites an output $a_{3}$ that is
behind a partial reflector $R_{2}$:
\begin{equation}
a_{3}=\sqrt{1-R_{2}^{2}}e^{-\alpha_{\textrm{z}}L_2/2}e^{i\varphi_2}a_{2}.\label{eq:a3}
\end{equation}
With the reflector in place,
the mode $a_{2}$ now also excites a left-traveling wave $c_{1}$, yielding
the following relation:
\begin{equation}
e^{-\alpha_{\textrm{z}}L_2/2 }e^{i\varphi_2}c_{1}=R_{2}e^{-\alpha_{\textrm{z}}L_2/2}e^{i\varphi_2}a_{2}.\label{eq:a3b}
\end{equation}
Both terms include the accumulated loss and an additional phase term~$\varphi_2$ accumulated while propagating over the length $L_2$.

The left-traveling wave $c_{1}$ excites a clockwise-propagating wave
$d_{2}$. The field $d_{2}$ propagates through the circumference
of the ring and excites another left-traveling wave $c_{2}$. The
interaction is described by a pair of equations that are analogous
to Eqs.~\ref{eq:b2}--\ref{eq:a2}:
\begin{eqnarray}
d_{2} & = & \sqrt{1-t^{2}}c_{1}+t\alpha e^{i\phi}d_{2}\label{eq:d2}\\
c_{2} & = & tc_{1}-\sqrt{1-t^{2}}\alpha e^{i\phi}d_{2}.\label{eq:c2}
\end{eqnarray}

The left-traveling wave $c_{2}$ reflects from the partial reflector
$R_{1}$ and contributes to input $a_{1}$, yielding:
\begin{equation}
a_{1}=\sqrt{1-R_{1}^{2}}e^{-\alpha_{\textrm{z}}L_1/2}e^{i\varphi_1}a_{0}-R_{1}e^{-\alpha_{\textrm{z}}L_1/2}e^{i\varphi_1}c_{2},\label{eq:a1}
\end{equation}
where $\varphi_1$ corresponds to the phase accumulated over the length $L_1$.

Equations~\ref{eq:a3}--\ref{eq:a1}, along with Eqs.~\ref{eq:b2}--\ref{eq:a2},
constitute the complete set of coupled equations. Solving for the
transmission yields our final equation:
\begin{eqnarray}
T(\phi) & = & \left|\frac{a_{3}}{a_{0}}\right|^{2}\nonumber \\
 & = & (1-R_{1}^{2})(1-R_{2}^{2})e^{-\alpha_{\textrm{z}}L}\times\nonumber \\
 &  & \;\left|\frac{(1-\alpha te^{i\phi})(t-\alpha e^{i\phi})}{(\alpha te^{i\phi}-1)^{2}-\gamma e^{i\varphi}(\alpha e^{i\phi}-t)^{2}}\right|^{2},\label{eq:asymmetric_ring}
\end{eqnarray}
with the substitutions:
\begin{eqnarray}
\gamma & \equiv & |R_{1}R_{2}e^{-\alpha_{\textrm{z}}L}|\label{eq:newgamma}\\
\varphi & \equiv & 2\pi(2L)n_{\textrm{eff}}/\lambda+\varphi_{0}.\label{eq:newvarphi}
\end{eqnarray}
The additional coefficients $\gamma$ and $\varphi$ correspond to
the  loss and phase contributed by the bus waveguide, respectively.
This result includes a new term, $e^{-\alpha_{\textrm{z}}L}$, accounting
for the propagation loss through the waveguide between the partial
reflectors, and a new pair of phase terms: $2\pi(2L)n_{\textrm{eff}}/\lambda$,
which accumulates along the length of the waveguide, and a phase offset $\varphi_{0}$,
which is introduced by the reflections. Equations~\ref{eq:newgamma}--\ref{eq:newvarphi} reveal that the strength and shape of the asymmetry only depend on the total round trip length $2L$, and not on the respective distance to the reflectors $L_1$ and $L_2$.

As expected, this result reduces to the ideal case (Eq.~\ref{eq:symmetric_ring_expanded}) when the
mirrors are removed ($R_{1},R_{2}\rightarrow0$). If we have no coupling
to the ring ($t\rightarrow1)$, we obtain the transmission for a Fabry-Perot
etalon of length $L$ with a pair of boundaries with reflectivity
$R_{1}$ and $R_{2}$~\cite{Yariv1991a},
\begin{equation}
T_{\textrm{Fabry-Perot}}(\varphi)=\left|\frac{\sqrt{1-R_{1}^{2}}\sqrt{1-R_{2}^{2}}e^{-\alpha_{\textrm{z}}L/2}}{1-R_{1}R_{2}e^{i\varphi}e^{-\alpha_{\textrm{z}}L}}\right|^{2}.\label{eq:Fabry_Perot}
\end{equation}

We scale Eq.~\ref{eq:asymmetric_ring} so that the peak value measured
at the output is 1. Doing so normalizes away any variables that are
not explicit functions of $\phi$ or $\varphi$, and yields:
\begin{eqnarray}
\tilde{T}(\phi) & = & \left|\frac{(1-\alpha te^{i\phi})(t-\alpha e^{i\phi})}{\gamma e^{i\varphi}(t-\alpha e^{i\phi})^{2}-(1-\alpha te^{i\phi})^{2}}\right|^{2}\nonumber \\
 & = & T_{\textrm{RR}}(\phi)\left|\frac{1}{1-\gamma e^{i\varphi}T_{\textrm{RR}}(\phi)}\right|^{2}.\label{eq:asymmetric_ring_correctionstyle}
\end{eqnarray}
Equation~\ref{eq:asymmetric_ring_correctionstyle} shows that $\tilde{T}(\phi)$
is just $T_{\textrm{RR}}(\phi)$ with a correction term tuned by the
asymmetry parameter $\gamma$.

We will show that this equation can be used to extract propagation
losses, even in the presence of strong boundary reflections.

\subsection{Asymmetry threshold}\label{SEC:Asymmetry_threshold}

If the reflections $R_{1}$ or $R_{2}$ are small and the bus waveguide
is very long, the asymmetry parameter $\gamma$ is correspondingly
small. One might expect the asymmetric correction to $T_{\textrm{RR}}(\phi)$
to become negligible. However, even small $\gamma$ values can lead
to pronounced asymmetries in some configurations. The term with $\gamma$
in the denominator dominates when
\begin{eqnarray}
1 & \ll & \left|\frac{2\gamma e^{i\varphi}(t-\alpha e^{i\phi})^{2}}{(1-\alpha te^{i\phi})^{2}}\right|.
\end{eqnarray}
This means that the reflection term $\gamma$ needs to be larger than
a threshold value of
\begin{equation}
\gamma_{\textrm{threshold}}\equiv\frac{(1-\alpha t)^{2}}{2|t+\alpha|^{2}}.\label{eq:threshold}
\end{equation}

We expect reflections (and thus $\gamma$) to be very small so as
not to observe any asymmetries. However, Eq.~\ref{eq:threshold}
shows that it is easy to obtain asymmetric resonances if both $\alpha$
and $t$ approach 1, as is generally desired. As was mentioned previously,
large values of $\alpha$ correspond to smaller rings or low propagation
losses.

For example, for a ring resonator with a diameter of \redline{100~$\upmu$m
and a propagation loss of 0.1~dB/mm, $\alpha\approx0.99$ (neglecting the insertion loss of the directional coupler). In this
case, $\gamma_{\textrm{threshold}}\approx10^{-5}$, and so just 0.25\%
($-\nobreak26$~dB)} reflections at the boundaries are necessary to cause noticeable asymmetries. \redline{A reflection this low is easily produced as a byproduct of a Y-junction splitter or a multimode interferometer, among other commonly used integrated photonic components.} Above
this threshold, the standard symmetric transfer function begins to
fail. Functionally, this means that asymmetries can be suppressed
in systems with considerable propagation loss or with large rings.
This low threshold also explains why asymmetries are so common in
high-$Q$ systems, and why it is hard to estimate the propagation
loss using rings in low-loss systems.

\section{Results and Discussion}

\subsection{Device geometry and simulation parameters}

We use a commercial 2-dimensional finite difference time domain (2D-FDTD)
solver and the effective index method~\cite{Hammer2009} to simulate
a micro-ring resonator in order to verify these equations. The device
we model consists of a lossy silicon ($n=3.5+0.0002i$) micro-racetrack
resonator on a silica ($n=1.45$) substrate. The racetrack is formed of waveguides
that are etched from a 220~nm thick silicon slab and have a horizontal width of 300~nm.
We set the constituent materials to be dispersionless to simplify the model and to
prove that these asymmetric artifacts are independent of any specific material dispersion.
The resonator parameters include a
coupling length of 10~$\upmu$m, an edge-to-edge coupling gap of
90~nm and a total ring circumference of 400~$\upmu$m. This device
is designed to achieve critical-coupling for TM-polarized light in
the center of the telecom operation range, at $\lambda=1550\textrm{ nm}$.
The imaginary component of the index is selected to achieve a propagation
loss near 5~dB/mm (Fig.~\ref{Fig: symmetric ring}a). Though this
loss might appear significant, we use a small ring for $\alpha$ to
remain large enough (\emph{$\alpha\approx0.75$)} so as not to suppress
the effects we are studying. Both of these factors contribute to minimizing
the simulation time. The complete set of parameters is summarized
in Table~\ref{tab:Simulation-parameters}.
\begin{table}[h]
\centering
\caption{Device geometry}
\label{tab:Simulation-parameters}
\begin{tabular}{|c|c|}
\hline
Material index & 3.5 + $2\times10^{-4}$i\tabularnewline
\hline
Substrate index & 1.45\tabularnewline
\hline
Waveguide width & 300 nm\tabularnewline
\hline
Waveguide height & 220 nm\tabularnewline
\hline
Ring circumference & 400 $\upmu$m\tabularnewline
\hline
Coupling length & 10 $\upmu$m\tabularnewline
\hline
Gap size & 90 nm\tabularnewline
\hline
\end{tabular}
\par
\end{table}

The effective index method yields an effective index of $n_{\textrm{eff}}=1.93$
for the waveguide and $n_{\textrm{eff}}=1$ for all other regions.
We calculate the propagation loss for TM-polarized light at $\lambda=1550\textrm{ nm}$
by performing a virtual cut-back method~\cite{Vlasov2004}; we monitor
the transmitted light at various positions in the waveguide and perform
a linear fit (Fig.~\ref{Fig: symmetric ring}a), yielding a propagation
loss of 5.2~dB/mm. For the given ring circumference and propagation
loss we calculate the loss coefficient at $\lambda=1550\textrm{ nm}$
to be \emph{$\alpha=0.775$}. The predicted $\alpha$ also includes
the minimal scattering loss that results from the mode mismatch between
the straight and bent parts of the racetrack resonator (0.2\% per
junction), \redline{but neglects the insertion loss of the directional coupler}.
Given the selected parameters, we estimate the maximum loaded $Q$-factor to be
$Q=\nicefrac{\pi n_{\textrm{g}}}{(\alpha_{\textrm{z}}\lambda)}=3.3\times10^{3}$~\cite{Preston2007}.

\subsection{Symmetric resonances}

\begin{figure}[htbp]
\centering\includegraphics[width=1\linewidth]{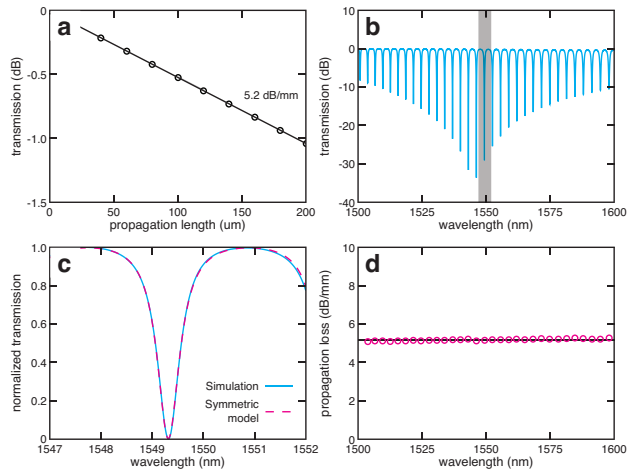}\caption{\textbf{a)} We obtain a propagation loss of 5.2~dB/mm for the waveguide
used in the resonator by fitting to the transmission of a straight
waveguide and extracting the slope. \textbf{b)} The transmission spectrum
for TM-polarized light in a micro-racetrack resonator with a total
circumference of 400~$\upmu$m. We observe near-critical coupling
throughout the wavelength range, peaking around $\lambda=1550\,\textrm{nm}$.
\textbf{c)} We fit to the resonance in the shaded region in \textbf{b
}and obtain $\alpha=0.774$. \textbf{d)} Extracted propagation loss
from each individual resonance in \textbf{b} (circles). The mean propagation
loss is 5.2~dB/mm, in agreement with the value fit in \textbf{a}. }
\label{Fig: symmetric ring}
\end{figure}

\begin{figure*}[htbp]
\centering\includegraphics[width=1\linewidth]{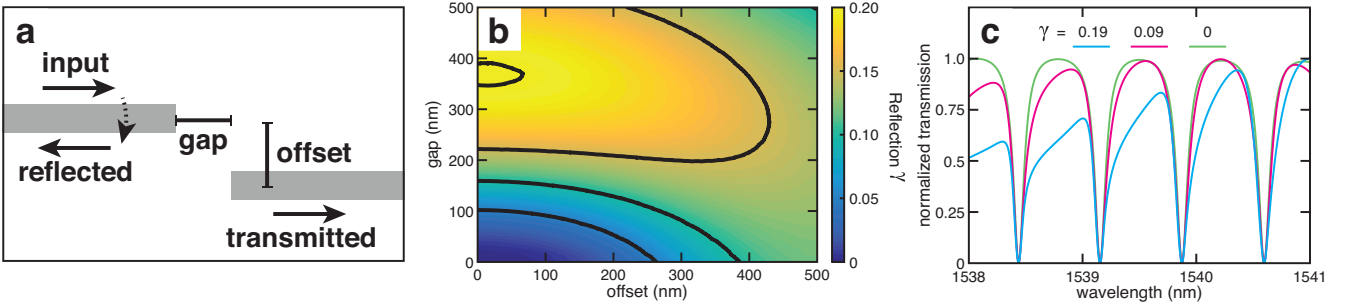}\caption{\textbf{a)} Top-down view of a pair of stitching errors in the waveguide surrounding
the ring resonator. They can behave as the partial reflectors in Fig.~\ref{Fig: ring_schematic}b
and cause asymmetries to form in the transmission spectrum of a ring
resonator. We describe these errors using two parameters, an offset
and a gap. \textbf{b)} Retrieved $\gamma$ (corresponding to $|R_{1}R_{2}|$)
for different offsets and gaps. \textbf{c)} As the reflections included
in the simulation become important, the asymmetries become more pronounced
without displacing the original resonances.}
\label{Fig:stitching_errors}
\end{figure*}

Before tackling the asymmetric case with paired partial reflections,
we begin by simulating an ideal ring resonator to demonstrate the
propagation loss extraction method~\cite{McKinnon2009}. The transmission
for TM-excitation in the geometry described above is shown in Fig.~\ref{Fig: symmetric ring}b.
We observe equally spaced symmetric resonance peaks with an average
free spectral range of 3.14~nm (390~GHz). They possess extinction ratios in excess of 10~dB throughout the
range of operation, confirming that the resonator is near-critically-coupled
to the waveguide. The largest extinction ratio observed (33~dB)
is located at $\ensuremath{\lambda=1546.2\textrm{~nm}}$, at the center
of the operation range. This resonance has a loaded $Q$-factor of
$3.0\times10^{3}$. The resonances ranging from $\lambda=1500$~to~$1600\textrm{~nm}$ are fit to the
traditional transfer function in Eq.~\ref{eq:symmetric_ring_expanded}, allowing the transmission coefficient $t$ and the total loss coefficient $\alpha$ to vary independently. A representative fit is shown in Fig.~\ref{Fig: symmetric ring}c. The fits are consistently well-behaved,
and yield a loss parameter $\alpha$ for each resonance. We calculate
the propagation losses $\alpha_{\textrm{z}}$ for each resonance using
Eq.~\ref{eq:alpha_loss} and plot them in Fig.~\ref{Fig: symmetric ring}d.
The loss values extracted using this method consistently agree well
with the propagation loss in Fig.~\ref{Fig: symmetric ring}a, with
a geometric mean of 5.2~dB/mm. This result demonstrates that fitting
to Eq.~\ref{eq:symmetric_ring_expanded} is a reliable method for
estimating the propagation loss of a waveguide for symmetric resonances.

\subsection{Asymmetric resonances\label{sub:Asymmetric-resonances}}

In order for asymmetries to appear in the simulated transmission spectrum,
partial reflectors must be placed into the waveguide.
We choose to insert artificial stitching errors to act as reflectors, as
they are simple to model and fully characterize using FDTD.
These errors routinely appear during the electron-beam lithography
process and are caused by the drift of successive write-windows, manifesting as horizontal displacements in the plane of the device layer (Fig.~\ref{Fig:stitching_errors}a). We simulate the reflection coefficient
for different gaps and offsets that might result from a stitching
error (Fig.~\ref{Fig:stitching_errors}b). For a perfectly aligned
pair of waveguides (\emph{i.e.,} no stitching error) none of the light
is reflected back towards the input, as expected. As the gap or offset
increases, the reflected field grows. Counter-intuitively, above a
certain gap size the reflected field decreases. This decrease is due
to interference effects from the combined field within the gap ---
the field reflected from the end of the input waveguide interferes
destructively with the field reflected from the beginning of the second
waveguide. Figure~\ref{Fig:stitching_errors}b shows that stitching
errors in our system yield a maximum reflection value of $\gamma=0.2$.
With $\alpha=0.775$, Eq.~\ref{eq:threshold} predicts that we need
$\gamma>\gamma_{\textrm{threshold}}=0.029$ before the transmission
spectrum from the simulated device exhibits significant asymmetries.

\begin{figure}[htbp]
\centering\includegraphics[width=1\linewidth]{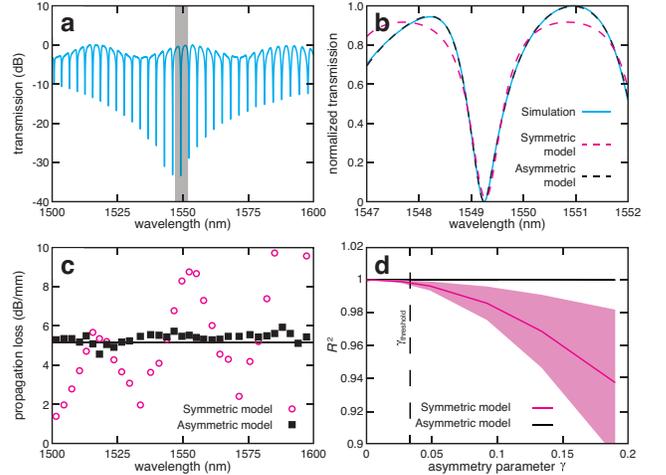}\caption{\textbf{a)} Transmission over the entire operation range for the device
described in Section~\ref{sub:Asymmetric-resonances}. \textbf{b)}
Fits to the resonance in the shaded area in \textbf{a}, using both
the standard symmetric model (red) and the asymmetric model developed
in this work (black). The symmetric model fits the center of the resonances
but is incapable of capturing the complete physics of the resonance.
\textbf{c)} The asymmetric model predicts a propagation loss of 5.4~dB/mm
while the symmetric model predicts an average propagation loss of
5.0~dB/mm. However, the standard symmetric model exhibits much larger
oscillating deviations. \textbf{d)} Shaded error bar plot of averaged
$R^{2}$ values of the fit to the transmission spectrum as a function
of asymmetry $\gamma$. The $R^{2}$ values remain constant for all
fits to the asymmetric model proposed in this work. On the other hand,\emph{
}the fit quality for the standard model declines once $\gamma>\gamma_{\textrm{threshold}}$. }
\label{Fig:asymmetric_rings}
\end{figure}

We set the partial reflectors in our device to be stitching gaps that
are 17~$\upmu$m apart, surrounding the ring-coupling region. This
length was chosen because it is not a multiple of the ring circumference
(400~$\upmu$m). If the spacing between the reflectors is much larger than the circumference of the ring, a ripple will form in the transfer function instead of an asymmetry. The elimination of this type of artifact has already been treated elsewhere in the literature~\cite{Strain2009}. We compare the effect of different asymmetry parameters
$\gamma$ by simulating devices with different stitching gap sizes
of 300~nm (corresponding to $\gamma=0.1902$), 200~nm ($\gamma=0.1339$),
150~nm ($\gamma=0.0921$), 100~nm ($\gamma=0.0484$), 70~nm ($\gamma=0.0259$)
and 35~nm ($\gamma=0.0053$). In Fig.~\ref{Fig:stitching_errors}c,
we plot the transmission for the most extreme case with a gap size
of 300~nm and the median case, with a gap size of 150~nm. We compare
them to the ideal case from the previous section. As expected, the
asymmetry increases with increasing $\gamma$. \redline{We have purposely chosen to implement unphysically large reflections in our simulation in order to compensate for the large propagation loss that was necessary for the convergence of the simulation. However, though undesired reflections in realistic integrated photonics circuits are likely to be smaller in physical implementations, pronounced asymmetries will still manifest so long as the propagation losses are low enough to satisfy the the asymmetry criteria outlined in Section~\mbox{\ref{SEC:Asymmetry_threshold}}.}

The complete transmission spectrum for the device with the largest
gap is shown in Fig.~\ref{Fig:asymmetric_rings}a. The familiar resonance
peaks from Fig.~\ref{Fig: symmetric ring}b are superimposed with
a slower secondary oscillation which adds characteristic asymmetries
to the individual resonances. The resonances now also appear to possess
different extinction ratios than their idealized counterparts. However,
this apparent difference is an artifact of the superposition between
the two oscillations.

We fit these resonances using both the standard symmetric transfer
function $T_{\textrm{RR}}(\phi)$, and the asymmetric transfer function
$\tilde{T}(\phi)$ (Fig.~\ref{Fig:asymmetric_rings}b). In simulation
we know the length of the bus waveguide between the reflectors; however,
if this length is unknown, its value could easily be determined
by taking the Fourier transform of the measured transmission. The
propagation loss extracted using both the symmetric and asymmetric
transfer functions are plotted in Fig.~\ref{Fig:asymmetric_rings}c.
The asymmetric function consistently predicts an average propagation
loss of 5.4~dB/mm with a geometric standard deviation of 1.05, in
agreement with the value extracted from the cut-back simulation and
the symmetric fits in Fig.~\ref{Fig: symmetric ring}. On the other
hand, the standard transfer function $T_{\textrm{RR}}(\phi)$ predicts
an average loss of 5.0~dB/mm with a geometric standard deviation
of 1.70. This large standard deviation is evident in the dramatic oscillations
about the mean, with values ranging from 1.4 to 11.6~dB/mm. The asymmetric
function fits the data significantly better when strong reflections
are present. To demonstrate this quantitatively, we calculated the
average $R^{2}$ for the fits in each spectrum. As can be seen in
Fig.~\ref{Fig:asymmetric_rings}d, both transfer functions fit the
data well when $\gamma<\gamma_{\textrm{threshold}}=0.029$, but above
this value the standard symmetric transfer function deviates significantly
from $R^{2}=1$ with a large range of $R^{2}$ values, whereas the
asymmetric transfer function remains at $R^{2}=1$ for every resonance.

\section{Conclusion}
We demonstrated that asymmetric resonances can result from purely
linear effects in micro-ring resonators, such as partial reflections
that are unavoidable in a realistic device or experimental setup.
We derived and numerically verified a threshold above which the asymmetries
become pronounced. Realistic devices easily exceed this threshold
and therefore exhibit asymmetric resonances. The asymmetries are most
pronounced in low-loss systems, where they appear for boundary reflections
on the order of 1\% or less. The reflectors do not need to be as close
to each other as they are in this work --- the asymmetries described
in this work can just as easily originate from waveguide boundaries
when the total waveguide length is comparable to the resonator length.

We derived a new transfer function that takes these asymmetries into
account. The equations in this model reduce to their symmetric counterparts
in the special case of negligible reflections. This asymmetric transfer
function outperforms the standard symmetric transfer function at modeling
resonances from a ring resonator when the reflections are significantly
above the threshold. The standard ring transfer function is therefore
unreliable at extracting propagation losses, and the new asymmetric
transfer function derived in this paper should be used in its place.
As material deposition and fabrication techniques improve, losses
will decrease and $Q$-factors will increase, exacerbating the prevalence
of asymmetries in state-of-the-art devices and thus the need for a
new model that can account for asymmetries.

\section*{Acknowledgements}
Several people contributed to the work described in this paper. OR conceived of the basic idea for this work, performed the simulations and analyzed the results. MGM helped perform the derivations and analysis. EM supervised the research and the development of the manuscript. The authors declare no competing financial interests. The authors acknowledge the help of Daryl I. Vulis and Philip Camayd-Mu\~{n}oz with FDTD and the help of Kelly Anne Miller with statistical analysis. OR wrote the first draft of the manuscript; all authors subsequently took part in the revision process and approved the final copy of the manuscript. Olivia Mello, Kelly Anne Miller and Dario Rosenstock provided feedback on the manuscript throughout its development. The authors thank Jeremy Upham and Sebastian A. Schulz for helpful discusisons. The research described in this paper was supported by the National Science Foundation under contracts ECCS-1201976 and PHY-1415236. OR acknowledges support from the Natural Sciences and Engineering Research Council of Canada.

\Urlmuskip=0mu plus 1mu\relax

\bibliographystyle{naturemagnourl}
\bibliography{library}

\end{document}